\documentclass[9pt,twocolumn,twoside]{osajnl}
\journal{ol}
\setboolean{shortarticle}{true} 

\usepackage[percent]{overpic}
\usepackage{bm, upgreek}

\usepackage[overlay,absolute]{textpos}
\newcommand\PlaceText[3]{%
	\begin{textblock*}{10in}(#1,#2)
		#3
	\end{textblock*}
}%

\newcommand{\mum}{$\upmu$m }
\newcommand{\mumN}{$\upmu$m}

\title{Watt-level dysprosium fiber laser at 3.15 $\upmu$m with 73\% slope efficiency}

\author[1,*]{R. I. Woodward}
\author[1]{M. R. Majewski}
\author[2]{G. Bharathan}
\author[2]{D. D. Hudson}
\author[2]{A. Fuerbach}
\author[1]{S. D. Jackson}

\affil[1]{Department of Engineering, Macquarie University, New South Wales, Australia}
\affil[2]{Department of Physics, Macquarie University, New South Wales, Australia}

\affil[*]{Corresponding author: robert.woodward@mq.edu.au}

\ociscodes{(140.3510) Lasers, fiber; (140.3070) Infrared and far-infrared lasers; (140.5680) Rare earth and transition metal solid-state lasers.}

 \doi{\url{http://dx.doi.org/10.1364/OL.43.001471}}

\begin{abstract}
Rare-earth-doped fiber lasers are emerging as promising high-power mid-infrared sources for the 2.6--3.0~\mum and 3.3--3.8~\mum regions based on erbium and holmium ions.
The intermediate wavelength range, however, remains vastly underserved, despite prospects for important manufacturing and defense applications.
Here, we demonstrate the potential of dysprosium-doped fiber to solve this problem, with a simple in-band-pumped grating-stabilized linear cavity generating up to 1.06~W at 3.15~\mumN.
A slope efficiency of 73\% with respect to launched power (77\% relative to absorbed power) is achieved---the highest value for any mid-infrared fiber laser to date.
Opportunities for further power and efficiency scaling are also discussed.
\end{abstract}

\setboolean{displaycopyright}{true}

\begin{document}

\maketitle

\PlaceText{25mm}{9mm}{Vol. 43, Issue 7, pp. 1471--1474 (2018); https://doi.org/10.1364/OL.43.001471}

Rare-earth-doped fluoride fibers are promising gain media to address the growing need for compact, high-brightness mid-infrared (mid-IR) laser sources, which will enable a range of new applications spanning medicine, advanced manufacturing and defense~\cite{Ebrahim-Zadeh2008}.
Requirements for these applications are varied, however, and it remains an active area of research to extend the parameter space of performance that can be delivered by mid-IR fiber laser technology.

Numerous sensing applications, for example, exploit characteristic absorption features of important functional groups and molecular fingerprints throughout the mid-IR, and thus require broad source bandwidths~\cite{Ebrahim-Zadeh2008}.
This is being enabled by the development of few-cycle high-\emph{peak}-power mid-IR fiber lasers~\cite{Tang2016,woodward_2017_70fs} to generate octave-spanning supercontinua through nonlinear phenomena~\cite{Hudson2017}.
For many other applications, however, high \emph{average} power monochromatic sources are required: e.g. laser countermeasures against heat-seeking missiles in the 3--5~\mum atmospheric transparency window and processing of important materials (e.g. polymers) which absorb strongly in the mid-IR~\cite{Ebrahim-Zadeh2008}.

\begin{figure}[bt]
	\includegraphics{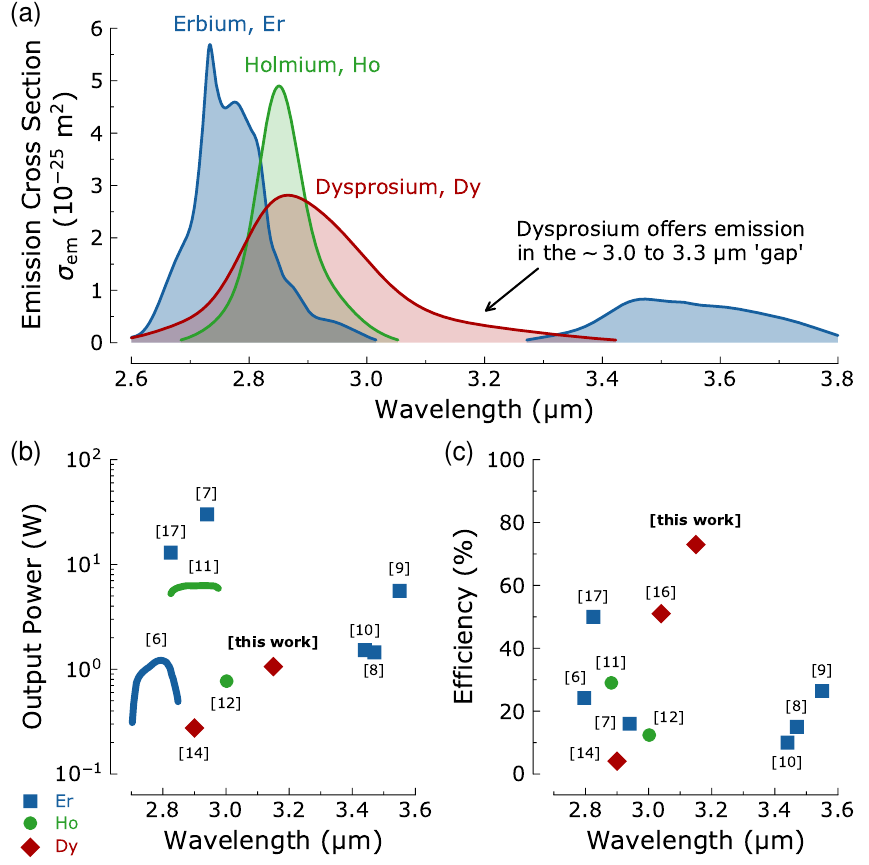}
	\caption{State-of-the-art in high-power mid-IR ZBLAN fiber lasers: (a) emission cross sections for available rare-earth dopants; (b) reported powers (lines indicate continuous tunability, references given in brackets); (c) reported efficiencies (slope efficiencies relative to launched power, apart from the Er 3.5~\mum transition, where optical-to-optical efficiencies are shown).}
	\label{fig:intro}
\end{figure}

To date, high-power fiber laser development has principally focused on erbium and holmium ions (Fig.~\ref{fig:intro}).
Er:ZBLAN has been shown to lase in the range 2.6--3.0~\mumN, including reports of watt-level tunability over more than 100~nm~\cite{Zhu2008, Liu2018} and up to 30~W output power at a fixed 2.94~\mum wavelength with 16\% slope efficiency~\cite{Fortin2015}.
By employing a dual-wavelength pump scheme, an additional erbium laser transition is accessible with 3.3--3.8~\mum emission~\cite{Henderson-Sapir2016a}, and up to 5.6~W has been generated at 3.55~\mum with 26\% efficiency (due to the dual-wavelength pump scheme, efficiency is better characterized by the optical-to-optical efficiency rather than slope efficiency)~\cite{Maes2017a, Fortin2016}.
For holmium, continuous lasing from $\sim$2.8--3.0~\mum at $>$7~W power has been reported (29\% slope efficiency)~\cite{Crawford2015, Li2011}.
Holmium also offers a 3.2~\mum laser transition, although this lies in-between high-energy manifolds and requires visible pump excitation; the prospects for high-performance are thus limited and to date, only 10~mW lasing output with 3\% efficiency has been achieved~\cite{Carbonnier1998}. 
Therefore, as Fig.~\ref{fig:intro} highlights, the 3.0--3.3~\mum region remains vastly underserved by current mid-IR sources.

Recently, dysprosium has emerged as a promising dopant for emission in this range~\cite{Jackson2003a}.
The transition from the first excited manifold to the ground state ($^6$H$_{13/2}$ $\rightarrow$ $^6$H$_{15/2}$) exhibits a broad emission cross section from 2.6 to 3.4~\mum [Fig.~\ref{fig:intro}(a)] and lasing with continuous tunability over nearly 600~nm in this range has been demonstrated~\cite{Majewski2018}.
The fact that this transition is between the lowest energy manifolds within the ion's energy level structure is also significant as it permits simple in-band pumping at 2.8~\mum~\cite{Majewski2016}.
This minimizes the quantum defect between pump and signal wavelengths, enabling higher slope efficiencies.
By contrast, 0.98~\mumN-pumped erbium and 1.15~\mumN-pumped holmium mid-IR lasers are fundamentally limited to maximum efficiencies of approximately 35\% and 40\%, respectively (unless advanced cavities with novel interplay between transitions are exploited: e.g. achieving 50\% in Er:ZBLAN~\cite{Aydin2017}).
The unused energy is dissipated as heat through multiphonon relaxation, which can be particularly damaging for high-power mid-IR lasers due to the poor thermo-mechanical stability of soft glasses compared to silicate fibers which are used in the near-IR.

Despite the promising spectroscopy of dysprosium, the highest power from a Dy:fiber laser to date is only 0.28~W (4.1\% slope efficiency)~\cite{Jackson2003a} and the best slope efficiency is 51\% (0.08~W maximum output)~\cite{Majewski2016}; both demonstrations were also free-running around the sub-3~\mum gain peak.
In this work, we advance the state-of-the-art in Dy:fiber lasers by demonstrating an optimized in-band-pumped fiber cavity, stabilized at 3.15~\mum by a custom-written fiber Bragg grating (FBG). 
Watt-level performance is achieved with 73\% slope efficiency, confirming the potential of dysprosium for power-scalable fiber lasers beyond 3~\mumN.

\begin{figure}[bt]
	\includegraphics{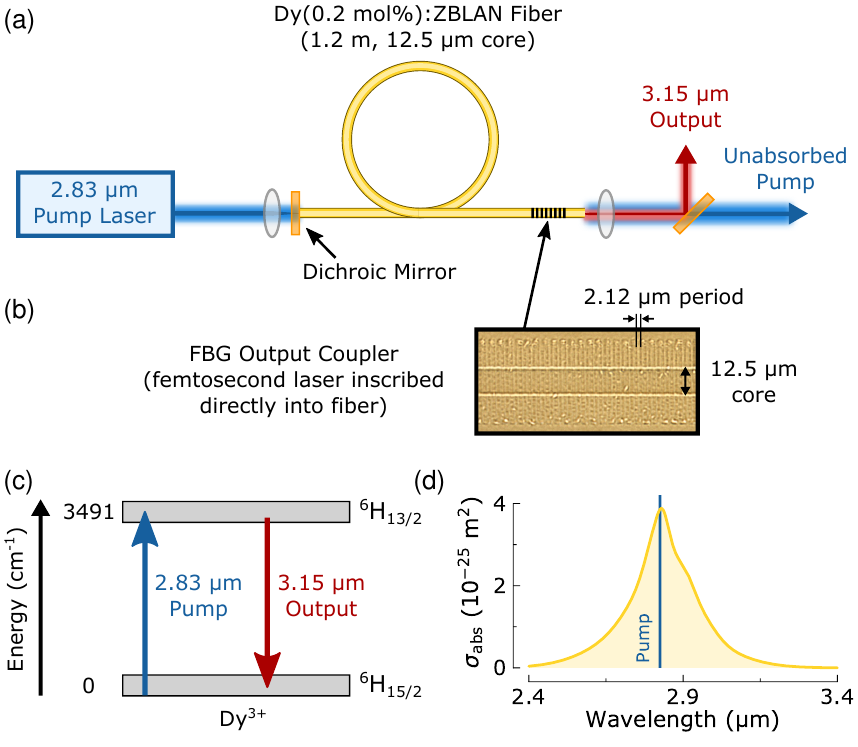}
	\caption{(a) Fiber laser cavity schematic. (b) Micrograph showing direct-written FBG structure in doped fiber core. (c) Transition energy levels in Dy$^{3+}$ (level positions from Ref.~\cite{Adam1988}). (d) Absorption cross section $\sigma_\mathrm{abs}$ of Dy:ZBLAN. }
	\label{fig:cavity}
\end{figure}

The laser setup comprises a simple linear cavity [Fig.~\ref{fig:cavity}(a)] with 1.2~m single-clad Dy(0.2~mol\%):ZBLAN fiber (Le Verre Fluor\'{e}).
The fiber has a core diameter of 12.5~\mum and 0.16 NA, yielding a single-mode cut-off wavelength of 2.6~\mum.
Therefore, the fiber is single-moded for both pump and signal wavelengths, maximizing modal overlap with the doped core.
A grating-stabilized Er:ZBLAN fiber laser is employed as the pump source at 2.825~\mum [corresponding to the peak of the dysprosium absorption cross section, Fig.~\ref{fig:cavity}(d)] with a high-brightness single-mode fiber output.
Pump light is collimated and refocused into the Dy:ZBLAN fiber using a pair of ZnSe aspheric lenses, with $\sim$80\% coupling efficiency.
Cavity feedback is provided by a butt-coupled CaF$_2$ dichroic mirror at the input fiber facet, which is highly reflective (>99\%) beyond 3~\mum and highly transmissive at the pump wavelength.
As the dysprosium emission cross-section peaks at $\sim$2.9~\mumN, spectrally selective feedback is required for wavelength-stabilized operation in the 3.0--3.3~\mum region of interest.
Therefore, a custom FBG is fabricated at the distal end to define the laser wavelength and act as the output coupler.

The FBG is directly inscribed into the active fiber using the core-scanned femtosecond-laser direct-write technique, similar to Ref.~\cite{Bharathan2017}.
Briefly, femtosecond laser pulses (800~nm, 115~fs, 145~nJ) are focused into the core of the Dy:ZBLAN fiber, which is mounted on a three-axis air-bearing translation stage.
By moving the fiber in a square-wave pattern transversely through the beam focus, planes of Type-I refractive index modifications are written into the fiber core, forming a FBG [Fig.~\ref{fig:cavity}(b)].
The spacing between planes defines the Bragg wavelength; here, we fabricate a second-order FBG with 2.12~\mum pitch over length 17~mm, yielding $\sim$60\% reflectivity at 3.15~\mumN.
After the FBG, the laser output is collimated and any remaining unabsorbed pump power is separated from the 3.15~\mum signal using a second dichroic mirror.

\begin{figure}[bt]
	\includegraphics{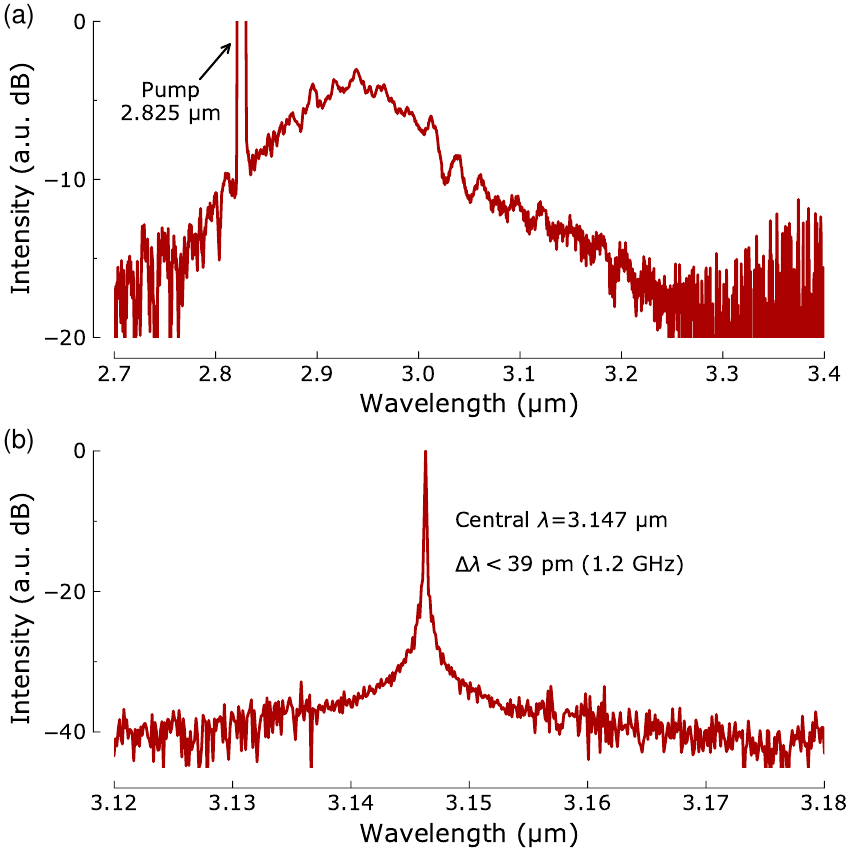}
	\caption{Optical spectra: (a) ASE output from in-band-pumped Dy fiber (with mirrors removed); (b) Laser spectrum.}
	\label{fig:spectrum}
\end{figure}

Pumping dysprosium at 2.825~\mum directly populates the first excited manifold $^6H_{13/2}$, with broad mid-IR emission from the transition back to the $^6H_{15/2}$ ground state [Fig.~\ref{fig:cavity}(c)].
The amplified spontaneous emission (ASE) output spectrum from the doped fiber is shown in Fig.~\ref{fig:spectrum}(a).
We note that this is narrower than the full emission cross section due to the nature of in-band pumping (the short-wavelength edge can be significantly enhanced with 1.7~\mum pumping of the higher energy $^6H_{11/2}$ manifold, albeit with a lower Stokes efficiency limit~\cite{Majewski2018}).

As the pump power is increased, lasing is observed at approximately 0.18~W threshold.
The output spectrum (measured with a Fourier transform interferometer) exhibits a stable single peak at 3.147~\mumN, in agreement with the target wavelength defined by our FBG [Fig.~\ref{fig:spectrum}(b)].
Measurement of the spectral width is limited by the $\sim$200~pm device resolution, although an improved resolution estimate of the linewidth can be achieved from the directly measured coherence length: i.e. the optical path length difference between interferometer arms where the amplitude of the interferogram envelope decreases by 1/$e$.
The coherence length was found to be longer than the physical delay length that is achievable on the interferometer (82~mm), i.e. $L_\mathrm{coh}>82$~mm, corresponding to an upper bound on laser linewidth of $\Delta f = c / (\pi L_\mathrm{coh}) = 1.2$~GHz (39~pm) assuming a Lorentzian-shaped spectrum.

\begin{figure}[bt]
	\includegraphics{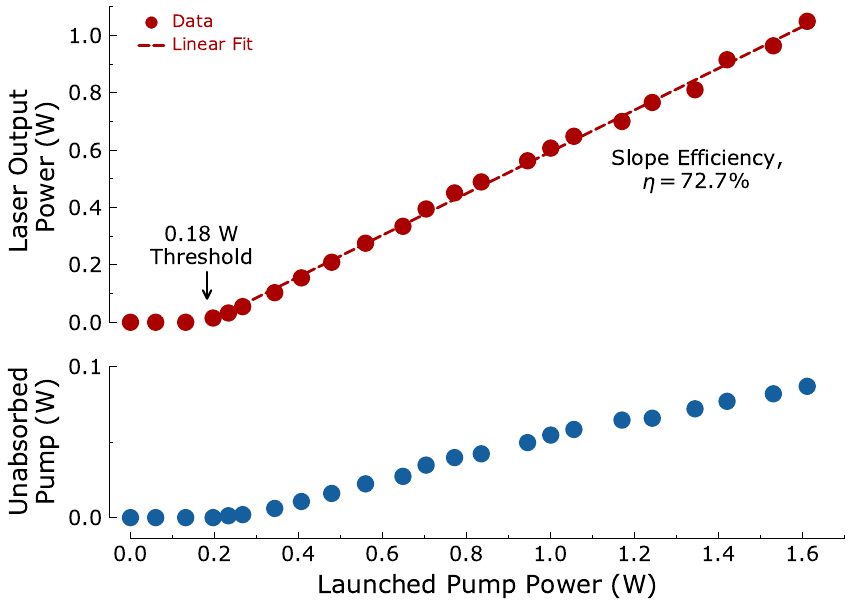}
	\caption{Laser output power and unabsorbed pump power relative to the launched pump power.}
	\label{fig:power}
\end{figure}

With increasing pump power, the laser output and unabsorbed pump power increase linearly (Fig.~\ref{fig:power}).
The strong linear fit to the data above threshold is in agreement with the fact that there are no known higher-order mechanisms involved in this laser transition which could cause deviation from this linearity (e.g. no excited state absorption, ESA---a feature which has plagued dysprosium lasers with other pump wavelengths~\cite{Gomes2010, Majewski2018}).
This is an additional benefit of dysprosium as a mid-IR laser ion: the transition is between the ground state and first excited level, avoiding the influence of complicated (and often deleterious) population kinetics which can arise when exciting higher lying levels.

Relative to the launched power, a slope efficiency of 72.7\% is recorded. 
Notably, this is the highest slope efficiency demonstrated to date from a mid-IR fiber laser, significantly improving upon previous leading results with up to 51\% efficiency~\cite{Majewski2016, Aydin2017}.
To assess the intrinsic efficiency of the gain fiber, we also compute the slope efficiency relative to absorbed pump power, by subtracting the measured residual pump from the launch power.
In this case, an efficiency of 77.0\% is recorded, highlighting the excellent potential for high efficiency mid-IR fiber lasers using dysprosium.

This slope efficiency is still less than the theoretical Stokes-limited efficiency of $\eta_\mathrm{max}=\lambda_\mathrm{signal}/\lambda_\mathrm{pump}=90$\%, however.
The primary reason for this is believed to be background loss in the fiber.
While fluoride fibers can theoretically offer even lower losses than silicates (due to lower Rayleigh scattering at longer wavelengths)~\cite{Shibata1981a}, state-of-the-art doped ZBLAN fiber fabrication still remains at the $\sim$0.1~dB/m attenuation level.
Measurement of loss in our Dy:ZBLAN fiber is complicated by the overlap of signal wavelength with the absorption cross section [Fig~\ref{fig:cavity}(d)].
Therefore, a cut-back loss measurement is performed at 3.39~\mum (effectively outside the absorption band), resulting in an attenuation value of $\sim$0.3~dB/m.
With ongoing improvements to fluoride fiber fabrication, we expect the achievable background loss values to decrease, leading to even higher slope efficiencies.

The maximum output power from our laser is 1.06~W.
At higher pump powers, the output power becomes highly unstable and decreases (the laser wavelength remains unchanged however, due to stabilization by the FBG).
While it has been shown that OH diffusion into ZBLAN fiber can cause absorption and heating that permanently degrades fiber tips, this is power and wavelength dependent~\cite{Caron2012}; near-instantaneous degradation is not expected for watt-level powers at 3.15~\mum as the wavelength is far from the OH bond resonance.
Additionally, we note that the fiber is polymer coated for protection and there is possibility of thermal stress through polymer absorption of the small fraction of unguided pump light.
However, we minimized this effect by removing a long section of coating near the input facet.
A more likely explanation for the power-limiting instability is the thermal and mechanical stresses at the soft-glass fiber input facet due to high-power core-guided pump and signal light: this affects the quality of the interface with the butt-coupled dichroic mirror, leading to reduced cavity feedback and thus, decreased laser performance.
This is supported by the observation that original laser performance is achieved again when reducing the pump power and re-optimizing the alignment (i.e. there did not appear to be permanent fiber damage).
Moving the cavity mirrors completely inside the fiber (e.g. using a second FBG) is expected to alleviate this problem, as recently demonstrated with a 5.6~W Er:ZBLAN fiber laser~\cite{Maes2017a}.
Further power scaling with Dy:ZBLAN fiber is thus a promising topic of future work to reach the multi-watt level.

\begin{figure}[bt]
	\includegraphics{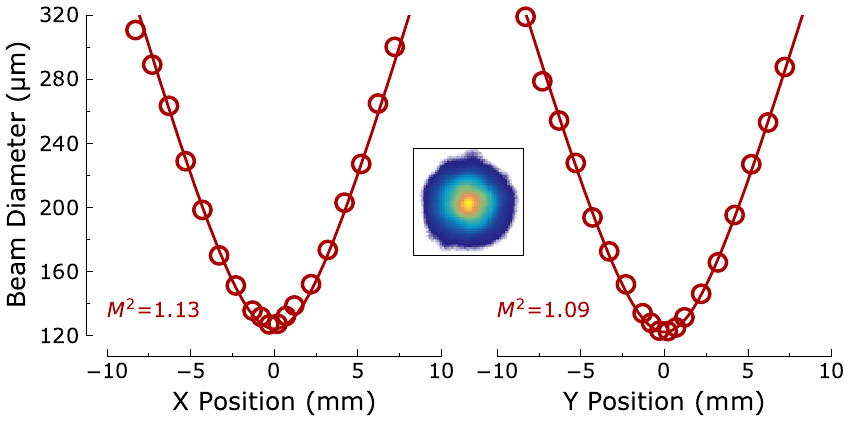}
	\caption{Beam quality measurement based on measured caustics in the $x$- and $y$-axis (inset: image of collimated beam).}
	\label{fig:m2}
\end{figure}

Laser beam quality is also characterized using a silicon microbolometer camera.
The collimated output from the fiber is focused using a 100~mm focal length CaF$_2$ lens, with the camera moved though the focus of the beam to measure the diameter (Fig.~\ref{fig:m2}).
The $M^2$ values for both axes are extracted to be $M^2<1.13$: close to the optimum diffraction-limited beam quality.

Before concluding, it should be noted that a number of alternative technologies also offer emission in the 3.0--3.3~\mum region. 
Optical parametric oscillators and single-pass parametric wavelength conversion schemes, for example, can deliver tunable multi-watt mid-IR output, although typically require a costly and complicated optical setup~\cite{Vainio2008, Murray2017}. 
Alternatively, quantum cascade lasers (QCLs) can generate light throughout the mid-IR with a minute footprint, although are fundamentally limited in terms of power scalability, particularly below 3.5~\mum where maximum powers of just 100s mW have been achieved to date~\cite{Razeghi2013}.
An additional interesting advance is the recent demonstration of 3.1~\mum gas lasers within hollow-core fibers at 1~W output power~\cite{Xu2017}, albeit with the added complexity of gas handling.
Rare-earth-doped fluoride fiber lasers, however, retain the unique advantage of combining a simple setup with high-brightness, power-scalable emission---offering to bring the well-established benefits of fiber lasers in the near-IR to the demanding mid-IR region.

In terms of our complete 3.15~\mum source, we note that by including the measured 20\% slope efficiency of the 2.83~\mum pump laser (diode-pumped at 0.98~\mumN) and the 50\% diode conversion efficiency, this gives an overall system wall plug efficiency (total electrical-to-optical efficiency) of $\sim$7\%, which compares favorably to the alternative mid-IR sources discussed above.
Additionally, by employing a doubly-resonant Er:ZBLAN pump laser design (with up to 50\% slope efficiency~\cite{Aydin2017}), this wall plug value could even reach 18\%.

In summary, we have reported the first watt-level fiber laser in the 3.0--3.3~\mum region, by employing dysprosium-doped ZBLAN fiber.
Using a simple in-band pumped linear cavity, 1.06~W at 3.15~\mum was generated with 73\% slope efficiency (relative to launched power): the highest efficiency for any mid-IR fiber laser to date.
This demonstration adds to the growing number of reports highlighting the promise of fluoride fiber lasers for mid-IR applications where a compact footprint, high brightness and high average power are important requirements.

\section*{Funding Information}
Australian Research Council (ARC) (DP140101336, DP170100531); Australian National Fabrication Facility (OptoFab Node, NCRIS).

\section*{Acknowledgment}
RIW acknowledges support through an MQ Research Fellowship.


\bigskip
\noindent

\begin{thebibliography}{10}
	\newcommand{\enquote}[1]{``#1''}
	
	\bibitem{Ebrahim-Zadeh2008}
	M.~Ebrahim-Zadeh and I.~T. Sorokina, \emph{{Mid-Infrared Coherent Sources and
			Applications}} (Springer, 2008).
	
	\bibitem{Tang2016}
	Y.~Tang, L.~G. Wright, K.~Charan, T.~Wang, C.~Xu, and F.~W. Wise,
	{\protect\JournalTitle{Optica}} \textbf{3}, 948 (2016).
	
	\bibitem{woodward_2017_70fs}
	R.~I. Woodward, D.~D. Hudson, A.~Fuerbach, and S.~D. Jackson,
	{\protect\JournalTitle{Opt. Lett.}} \textbf{42}, 4893 (2017).
	
	\bibitem{Hudson2017}
	D.~D. Hudson, S.~Antipov, L.~Li, I.~Alamgir, T.~Hu, M.~E. Amraoui,
	Y.~Messaddeq, M.~Rochette, S.~D. Jackson, and A.~Fuerbach,
	{\protect\JournalTitle{Optica}} \textbf{4}, 1163 (2017).
	
	\bibitem{Zhu2008}
	X.~Zhu and R.~Jain, {\protect\JournalTitle{IEEE Photon. Technol. Lett.}}
	\textbf{20}, 156 (2008).
	
	\bibitem{Liu2018}
	J.~Liu, M.~Wu, B.~Huang, P.~Tang, C.~Zhao, D.~Shen, D.~Fan, and S.~K. Turitsyn,
	{\protect\JournalTitle{IEEE J. Sel. Top. Quantum Electron.}} \textbf{24}
	(2018).
	
	\bibitem{Fortin2015}
	V.~Fortin, M.~Bernier, S.~T. Bah, and R.~Vallee, {\protect\JournalTitle{Opt.
			Lett.}} \textbf{40}, 2882 (2015).
	
	\bibitem{Henderson-Sapir2016a}
	O.~Henderson-Sapir, S.~D. Jackson, and D.~J. Ottaway,
	{\protect\JournalTitle{Opt. Lett.}} \textbf{41}, 1676 (2016).
	
	\bibitem{Maes2017a}
	F.~Maes, V.~Fortin, M.~Bernier, and R.~Vall{\'{e}}e,
	{\protect\JournalTitle{Opt. Lett.}} \textbf{42}, 2054 (2017).
	
	\bibitem{Fortin2016}
	V.~Fortin, F.~Maes, M.~Bernier, S.~T. Bah, M.~D'Auteuil, and R.~Vall{\'{e}}e,
	{\protect\JournalTitle{Opt. Lett.}} \textbf{41}, 559 (2016).
	
	\bibitem{Crawford2015}
	S.~Crawford, D.~D. Hudson, and S.~D. Jackson, {\protect\JournalTitle{IEEE
			Photonics J.}} \textbf{7}, 1 (2015).
	
	\bibitem{Li2011}
	J.~Li, D.~D. Hudson, and S.~D. Jackson, {\protect\JournalTitle{Opt. Lett.}}
	\textbf{36}, 3642 (2011).
	
	\bibitem{Carbonnier1998}
	C.~Carbonnier, H.~Tobben, and U.~B. Unrau, {\protect\JournalTitle{Electron.
			Lett.}} \textbf{34}, 893 (1998).
	
	\bibitem{Jackson2003a}
	S.~D. Jackson, {\protect\JournalTitle{Appl. Phys. Lett.}} \textbf{83}, 1316
	(2003).
	
	\bibitem{Majewski2018}
	M.~R. Majewski, R.~I. Woodward, and S.~D. Jackson, {\protect\JournalTitle{Opt.
			Lett.}} \textbf{43}, 971 (2018).
	
	\bibitem{Majewski2016}
	M.~R. Majewski and S.~D. Jackson, {\protect\JournalTitle{Opt. Lett.}}
	\textbf{41}, 2173 (2016).
	
	\bibitem{Aydin2017}
	Y.~O. Aydin, V.~Fortin, F.~Maes, F.~Jobin, S.~D. Jackson, R.~Vall{\'{e}}e, and
	M.~Bernier, {\protect\JournalTitle{Optica}} \textbf{4}, 235 (2017).
	
	\bibitem{Bharathan2017}
	G.~Bharathan, R.~I. Woodward, M.~Ams, D.~D. Hudson, S.~D. Jackson, and
	A.~Fuerbach, {\protect\JournalTitle{Opt. Express}} \textbf{25}, 30013 (2017).
	
	\bibitem{Adam1988}
	J.~L. Adam, A.~D. Docq, and J.~Lucas, {\protect\JournalTitle{J. Solid State
			Chem.}} \textbf{75}, 403 (1988).
	
	\bibitem{Gomes2010}
	L.~L. Gomes, A.~F. H. A. F.~H. Librantz, and S.~D. Jackson,
	{\protect\JournalTitle{J. Appl. Phys.}} \textbf{107}, 053103 (2010).
	
	\bibitem{Shibata1981a}
	S.~Shibata, M.~Horiguchi, K.~Jinguji, S.~Mitachi, T.~Kanamori, and T.~Manabe,
	{\protect\JournalTitle{Electron. Lett.}} \textbf{17}, 775 (1981).
	
	\bibitem{Caron2012}
	N.~Caron, M.~Bernier, D.~Faucher, and R.~Vall{\'{e}}e,
	{\protect\JournalTitle{Opt. Express}} \textbf{20}, 22188 (2012).
	
	\bibitem{Vainio2008}
	M.~Vainio, J.~Peltola, S.~Persijn, F.~J.~M. Harren, and L.~Halonen,
	{\protect\JournalTitle{Opt. Express}} \textbf{16}, 11141 (2008).
	
	\bibitem{Murray2017}
	R.~T. Murray, T.~H. Runcorn, S.~Guha, and J.~R. Taylor,
	{\protect\JournalTitle{Opt. Express}} \textbf{25}, 6421 (2017).
	
	\bibitem{Razeghi2013}
	M.~Razeghi, N.~Bandyopadhyay, Y.~Bai, Q.~Lu, and S.~Slivken,
	{\protect\JournalTitle{Opt. Mater. Express}} \textbf{3}, 1872 (2013).
	
	\bibitem{Xu2017}
	M.~Xu, F.~Yu, and J.~Knight, {\protect\JournalTitle{Opt. Lett.}} \textbf{42},
	4055 (2017).
	
\end{thebibliography}
\end{document}